\documentclass[pre,aps,floatfix,twocolumn,showpacs,preprintnumbers,superscriptaddress,showkeys]{revtex4-1}
\usepackage{graphicx}
\usepackage{amsmath}
\usepackage{amssymb}
\usepackage{color}

\def\w{\omega}
\def\W{\Omega}

\def\e{\varepsilon}
\def\vp{\varphi}
\newcommand{\ii}{\mathrm{i}}

\newcommand{\ee}{\mathrm{e}}

\newcommand{\const}{\mbox{const}}

\newcommand{\ds}{\displaystyle}

\renewcommand{\le}{\leqslant}
\renewcommand{\ge}{\geqslant}

\begin{document}
\title{Two types of quasiperiodic partial synchrony in oscillator ensembles}
\author{Michael Rosenblum}
\affiliation{Department of Physics and Astronomy, University of Potsdam,
 Karl-Liebknecht-Str. 24/25, D-14476 Potsdam-Golm, Germany}
\author{Arkady Pikovsky}
\affiliation{Department of Physics and Astronomy, University of Potsdam,
 Karl-Liebknecht-Str. 24/25, D-14476 Potsdam-Golm, Germany}
  \affiliation{Department of Control Theory, Nizhni Novgorod State University,
  Gagarin Av. 23, 606950, Nizhni Novgorod, Russia}
\date{\today}

\begin{abstract}
We analyze quasiperiodic partially synchronous states in an ensemble
of Stuart-Landau oscillators with global nonlinear coupling.
We reveal two types of such dynamics: in the first case the time-averaged 
frequencies of oscillators and of the mean field differ, 
while in the second case they are equal, but the motion of oscillators
is additionally modulated. We describe transitions from the synchronous state 
to both types of quasiperiodic dynamics, and a
transition between two different quasiperiodic states. 
We present an example of a bifurcation diagram,
where we show the borderlines for all these transitions, as well as 
domain of bistability.  
\end{abstract}

\pacs{05.45.Xt, %
05.65.+b%
}
\keywords{Oscillator populations, collective dynamics, quasiperiodic dynamics}

\maketitle

\section{Introduction}

An ensemble of globally coupled limit-cycle oscillators is a widely used model 
for many natural systems 
\cite{Winfree-67,*Winfree-80,Kuramoto-84,Strogatz-00,*Strogatz-03,%
Pikovsky-Rosenblum-Kurths-01,Golomb-Hansel-Mato-01,*Ott-book-02,%
*Manrubia-Mikhailov-Zanette-04,*Acebron-etal-05,%
*Breakspear-Heitmann-Daffertshofer-10}.
The main, well-understood, effect in this setup is emergence of a 
collective mode (mean field) via synchronization of ensemble elements
\cite{Kuramoto-75,*Sakaguchi-Kuramoto-86,Kuramoto-84}.
Typical synchronization scenario is as follows.
If a homogeneous coupling between generally non-identical ensemble units
is attractive and 
quantified by parameter $\e>0$,
then, typically, with the increase of $\e$ beyond some critical value, 
a number of oscillators adjust their (initially different) frequencies and form 
a synchronous group.
The units in this group have coherent, though slightly different, phases
and, as a result, produce a non-zero mean field. 
With the further increase of $\e$, more and more oscillators join the 
synchronous group and the mean field amplitude grows.
The situation becomes almost trivial if the oscillators in the ensemble 
are \textit{identical}: then full synchrony appears already for an arbitrarily 
small attractive coupling.

The described synchronization scenario assumes that the interaction 
remains attractive for all values of $\e$ and for all amplitudes 
of the collective mode.  This is, however,  not a general case:
one can expect that weak and strong forcing on an oscillator may have 
different properties.
In this paper we are interested exactly in the situations when the 
increase of the bifurcation parameter $\e$ results in a change of the interaction 
type from an attractive to a repulsive one, yielding complex regimes already for 
the simplest setups with identical units. 
In particular, our goal is to analyze
the \textit{quasiperiodic partially synchronous (QPS) regimes} that appear via 
a synchrony-breaking transition and are characterized by 
scattered or clustered oscillator states and yet non-vanishing collective mode;
the most important feature is the quasiperiodic dynamics of oscillators. 
We describe two types of such solutions. 
In one case, which we denote as QPS-I, the time averaged frequencies of individual 
oscillators and of the mean field are different. 
In the second case, labeled as QPS-II,
the averaged frequencies coincide, but the motion of oscillators is 
additionally modulated.
We demonstrate that these regimes between full synchrony and complete 
asynchrony appear 
in an ensemble of Stuart-Landau (SL) oscillators with global nonlinear coupling.
For this model, we analytically find the conditions for two bifurcations 
resulting in emergence of two types of QPS dynamics. Furthermore, we reveal 
transitions between the QPS-I and the QPS-II dynamics, as well as parameter domains 
where full synchrony coexists with the QPS states.

The paper is organized as follows. First we introduce our basic model in 
Section \ref{NlSL}. Then, in Section \ref{weak_lim} we discuss the weak-coupling 
limit and properties of partial synchrony in the phase approximation. 
Next, in Section \ref{sec:bpa} we analyze stability of the synchronous and of
the asynchronous state for arbitrary coupling, and present the diagram of 
different states.
Section \ref{sec:num} presents the results of numerical analysis.
We discuss and summarize our results in Section \ref{SumDis}.

\section{Stuart-Landau oscillators with global nonlinear coupling}
\label{NlSL}

Our basic model is a system of $N$ identical Stuart-Landau oscillators 
with global nonlinear coupling 
(cf.~\cite{Pikovsky-Rosenblum-09,PhysRevLett.94.164102,Schmidt_etal-14,*PhysRevLett.114.034101}):
\begin{align}
\dot a_k=(1+\ii\w_0)&a_k-(1+\ii\alpha)|a_k|^2a_k \nonumber \\
&+(\e_1+\ii\e_2)A-(\eta_1+\ii\eta_2)|A|^2A \;,
\label{e1}
\end{align}
where
\begin{equation}
A=\rho \ee^{\ii\Theta}=N^{-1}\sum_{k=1}^N a_k \;
\label{nk3}
\end{equation}
is the complex mean field.
Here $\w_0$ is the frequency of small oscillations 
(it does not play any role since it can be eliminated by a transformation to
a rotating reference frame)
and $\alpha$ describes non-isochronicity of uncoupled oscillators.
The coupling is quantified by four parameters: parameters $\e_{1,2}$ describe linear coupling term
$\sim A$,
while parameters  $\eta_{1,2}$ describe nonlinear coupling $\sim |A|^2A$.  
Notice that the case of purely linear coupling was extensively studied by 
Hakim and Rappel \cite{PhysRevA.46.R7347}
and by Nakagawa and Kuramoto \cite{Nakagawa-Kuramoto-93,*Nakagawa-Kuramoto-94},
see also \cite{Matthews1991293,*PhysRevLett.93.104101,*PhysRevLett.96.054101}. 
We come back to this case in the discussion section below. 

In the synchronous regime 
$a_1=\ldots=a_N=r\ee^{\ii\vp}=A$ and the stationary 
(uniformly rotating with frequency $\W$)
solution of Eqs.~(\ref{e1}) can be easily found:
\begin{equation}
r^2=\frac{1+\e_1}{1+\eta_1}\;,\qquad
\W=\dot\vp=\w_0+\e_2-\frac{(\alpha+\eta_2)(1+\e_1)}{1+\eta_1}\;.
\label{e1_1}
\end{equation}
In the fully asynchronous regime the mean field $A$ vanishes. 
This state is microscopically degenerate, as
there is just one condition on the distribution of $N$ phases. 
Stability of the asynchronous and synchronous states is studied in 
the next sections.

Before proceeding to a more general analysis, we mention one intermediate 
dynamical state which appears for a special set of parameters. 
Indeed, if the ratio $\frac{\e_1+\ii\e_2}{\eta_1+\ii\eta_2}$ is real,
then for $|A|^2=\frac{\e_1+\ii\e_2}{\eta_1+\ii\eta_2}$ the coupling vanishes. 
In this regime, called ``bunch state'' in~\cite{Pikovsky-Rosenblum-09}, 
the oscillators are partially synchronized, but the dynamics
is purely periodic as all units have the same frequency $\w_0-\alpha$.

\section{Weak coupling limit, phase approximation, and partial synchrony}
\label{weak_lim}

\subsection{Weak coupling limit}
\label{wcl}
Close to the asynchronous regime, where the amplitude $A$ is small, the coupling between
the oscillators is weak. This also holds for a non-small amplitude $A$, if 
coupling parameters are small $\e_{1,2}\ll 1$, $\eta_{1,2}\ll 1$.
For such a weak coupling,
 the amplitudes $|a_k|$ are only slightly perturbed: $|a_k|\approx 1$.
Then $a_k\approx \ee^{\ii\vp_k}$ 
and the mean field $A$ is
simply the Kuramoto order parameter $R\exp[\ii\Theta]=N^{-1}\sum_j\exp[\ii\vp_j]$, 
  $0\le R\le 1$.
Using the standard approach \cite{Kuramoto-84,Pikovsky-Rosenblum-Kurths-01},
Eq.~(\ref{e1}) can be reduced to the phase dynamics:
\begin{equation}
 \dot\vp_k 
=\overline{\w}+{\cal E}(R;\e_{1,2},\eta_{1,2})R\sin\left [
\Theta-\vp_k+\beta(R;\e_{1,2},\eta_{1,2})\right ]\;,
\label{e1_2}
\end{equation}
where $\overline{\w}=\w_0-\alpha$, 
while the amplitude ${\cal E}\ge 0$ and the phase shift $\beta$ in the coupling
are determined from
\begin{equation}
\begin{aligned}
 {\cal E}\sin\beta&=\alpha(R^2 \eta_1-\e_1)+\e_2- R^2\eta_2\;,\\
  {\cal E}\cos\beta&=\e_1-R^2\eta_1-\alpha(R^2 \eta_2-\e_2)\;.
\end{aligned}
\label{eq:ebeta}
\end{equation}

Equation~(\ref{e1_2}) can be considered as a nonlinear generalization of the
popular Kuramoto-Sakaguchi model \cite{Kuramoto-75,*Sakaguchi-Kuramoto-86}.
Indeed, for the linear coupling $\eta_1=\eta_2=0$ we obtain exactly
the Kuramoto-Sakaguchi model
\begin{equation}
 \dot\vp_k =\overline{\w}+\e R\sin(\Theta-\vp_k+\beta)\;,
\label{eq:ks}
\end{equation}
where 
\[
 \e^2=(\alpha\e_1-\e_2)^2+(\alpha\e_2+\e_1)^2\;, \;
\beta=\text{Arg}(\e_1+\alpha\e_2+\ii(\e_2-\alpha\e_1))\;.
\]
Generally, if both linear and nonlinear couplings are present,
instead of two constants $\e$ and $\beta$
we have two functions 
${\cal E}(R;\e_{1,2},\eta_{1,2}),\;\beta(R;\e_{1,2},\eta_{1,2})$; 
this model has been suggested and studied 
in \cite{Rosenblum-Pikovsky-07,Pikovsky-Rosenblum-09}. 
Notice that this phase model also appears 
as an approximation of the system of SL oscillators,
coupled through a common nonlinear load~\cite{Pikovsky-Rosenblum-09}.
A very important property of the model (\ref{e1_2}) is its partial
integrability: according to the Watanabe-Strogatz 
theory~\cite{Watanabe-Strogatz-93,*Watanabe-Strogatz-94,*Pikovsky-Rosenblum-11}, 
dynamics of (\ref{e1_2}) is described by three global variables 
and $N-3$ constants of motion; this description is valid
for any $N>3$, including the thermodynamic limit $N\to\infty$,
  
It is easy to see that synchronous solution of the model 
(\ref{e1_2}) is stable if \\
${\cal E}(1;\e_{1,2},\eta_{1,2})\cos\left[
\beta(1;\e_{1,2},\eta_{1,2})\right ]<0$.
To determine stability of the asynchronous state, we have 
to consider effect of a small perturbation, i.e. effect 
of the mean field with $R\ll1$. This means that we can neglect
the terms $R^2\eta_{1,2}$ in~\eqref{eq:ebeta} and the model
reduces to the Kuramoto-Sakaguchi system (\ref{eq:ks}).
So, only the linear part of the coupling contributes
to the instability of the asynchronous state.
The asynchronous state will be unstable if the coupling is attractive, 
i.e. if $\cos\beta>0$. This condition yields instability
provided $\e_1+\alpha\e_2>0$, and stability otherwise. 

\subsection{Partial synchrony and quasiperiodicity within phase approximation}
\label{sec:soq}

Here we discuss partial synchrony in the framework of the
phase approximation~\eqref{e1_2}.
A detailed analysis of this model has been 
presented in~\cite{Rosenblum-Pikovsky-07,Pikovsky-Rosenblum-09}, so we just
reproduce the basic ideas for consistency.

Consider first the pure Kuramoto-Sakaguchi case~\eqref{eq:ks}. 
As is well-known, the synchronous state, 
$R=1$, is stable, if $|\beta|<\pi/2$, and unstable, if $\pi/2<\beta<3\pi/2$ 
(we remind that $\e>0$).
For the asynchronous (splay) state,  $R=0$, the stability conditions are reversed.
Hence, there occurs either full synchrony or full asynchrony. 
Notice that existence of other attractive states with $0<R<1$ or 
of many-cluster solutions with $R=0$ 
is excluded by the Watanabe-Strogatz theory 
\cite{Watanabe-Strogatz-93,*Watanabe-Strogatz-94,*Pikovsky-Rosenblum-11}.

Complementarity of stability domains for synchronous and asynchronous solutions 
is a specific property of the Kuramoto-Sakaguchi model. 
For general globally coupled systems the situation can be different. 
So, we can expect overlap of stability domains for some parameter region; 
then, in this region, the system is at least bi-stable. 
Another possible case, of our interest here, is when \textit{both fully synchronous 
and fully asynchronous solutions are unstable}. 
Then, for the corresponding parameters the system is enforced to settle at some 
non-trivial state between synchrony and asynchrony. 
An example is given 
by Eq.~(\ref{e1_2}), where $\beta(R;\e_{1,2},\eta_{1,2})$ depends on 
the order parameter $R$. Stability 
of the asynchronous state is determined solely by $\beta(0,\e_{1,2},\eta_{1,2})$, 
while for the state of 
full synchrony, $R=1$, the value $\beta(1,\e_{1,2},\eta_{1,2})$
is relevant. If $\cos\beta(0,\e_{1,2},\eta_{1,2})>0$ and $\cos\beta(1,\e_{1,2},\eta_{1,2})<0$,
both fully synchronous and asynchronous states are unstable (the border between
stability and instability domains for the synchronous state is determined 
from the condition $\cos\beta(1,\e_{1,2},\eta_{1,2})=0$).
It means, that an intermediate, \textit{partially synchronous}
state with $0< R< 1$ is established. 
The order parameter in this state is given by the 
condition $\cos\beta(R,\e_{1,2},\eta_{1,2})=0$. 

Next, we stress that system (\ref{e1_2}), like the Kuramoto-Sakaguchi model, 
is fully described by the Watanabe-Strogatz 
theory~\cite{Watanabe-Strogatz-93,*Watanabe-Strogatz-94,*Pikovsky-Rosenblum-11} 
which excludes the states
with more than one synchronous cluster for general non-identical initial 
conditions. Hence, at partial synchrony, all phases shall be scattered and 
non-uniformly distributed on the unit circle. 
As it follows from Eq.~(\ref{e1_2}), this scattering results in 
different  instantaneous frequencies of all units.
Furthermore, it results in the most peculiar feature 
of this state, namely in a difference of the time-average 
frequencies of the units and of the frequency of the mean field. Let us denote these
frequencies as $\W$ and $\nu$, respectively. 
(Notice that since oscillators are identical, all $\W_k=\W$).
In our previous publications \cite{Rosenblum-Pikovsky-07,Pikovsky-Rosenblum-09}
we called such states with $0<R<1$ and $\W\ne\nu$  
self-organized quasiperiodic (SOQ) solutions.

Qualitatively the property $\nu\ne\W=\langle \dot\vp \rangle$ 
can be shown by contradiction. Suppose first the contrary, $\nu=\W$,
and consider the motion in the frame, rotating with the 
mean field frequency $\nu$.  
Then, according to Eq.~(\ref{e1_2}), the points, representing 
some oscillators move forwards
and some of them move backwards with respect to the mean field.
Hence, there are two values of $\vp$ where the velocity 
in this frame changes its sign, and one of these values 
corresponds to stable state and another corresponds to an 
unstable one. 
So, the oscillators having 
these phase values are in rest. Other oscillators move 
towards the stable state and therefore merge
into a cluster. Since clusters in this setup are not possible, 
the assumption $\nu=\W$ cannot be true. Hence, either all oscillators 
move faster than the mean field or all of them move slower, 
i.e. $\W\ne\nu$.
A detailed quantitative analysis of system (\ref{e1_2}) can be 
found in \cite{Rosenblum-Pikovsky-07,Pikovsky-Rosenblum-09}.

It is important to notice that the phase model~\eqref{e1_2} is only 
an approximation of the full system of Eqs.~\eqref{e1} for the case 
when the amplitude dynamics is enslaved. 
In this situation the amplitude perturbations decay rapidly, and
instability of the fully synchronous state occurs due to one real 
eigenvalue, corresponding to the phase (as described in the next section).
Thus, for weak coupling we can expect that 
the above described SOQ dynamics
appears close to instability of the synchronous state of the full system,
when one real multiplier becomes larger than unity. 
Here we denote such dynamics as QPS-I, to be 
distinguished from another quasiperiodic state, discussed below. However, 
the correspondence between QPS-I and SOQ is not exact, as due to corrections
to the first-order model~\eqref{e1_2}, some fine features may become different.
For example, while in model~\eqref{e1_2} several clusters are not possible due to 
the Watanabe-Strogatz theory, 
already small perturbations to the model generally destroy this property and 
enable clustering.

\section{Beyond the phase approximation}
\label{sec:bpa}

We analyze stability of the fully synchronous state Eq.~(\ref{e1_1}) with respect 
to the evaporation of individual oscillators from the synchronous cluster. 
In fact, one can always consider purely transversal evaporation modes such that the  mean field $A$ 
remains unchanged. 
Substituting $a_k=b_ke^{i\W t}$, we make transformation to the 
coordinate frame, rotating with the frequency $\W$, 
where $\W$ is given by Eq.~(\ref{e1_1}). 
We obtain
\begin{align}
\dot b_k=(1+\ii\w)&b_k-(1+\ii\alpha)|b_k|^2b_k \nonumber  \\
& + (\e_1+\ii\e_2)B-(\eta_1+\ii\eta_2)|B|^2B \;,
\label{e2}
\end{align}
where $\w=\w_0-\W$ and $B=N^{-1}\sum_j^N b_j$.
In the new frame, synchronous motion corresponds to a resting point; 
we choose the coordinate system so that $b_k=B=r$.
Linearizing the equation around this point while keeping $B=\const$, 
we obtain after straightforward manipulations the eigenvalues
\begin{equation}
\lambda_{1,2}=(1-2r^2)\pm\sqrt{(1-3\alpha^2)r^4+4\w\alpha r^2-\w^2}\;,
\label{eigenv}
\end{equation}
related to evaporation 
multipliers as $|\mu|=\ee^{\lambda T}$, where $T=2\pi/\W$ is the oscillation 
period~\cite{Kaneko-94,*Pikovsky-Popovych-Maistrenko-01,Rosenblum-Pikovsky-07}.

If both eigenvalues~\eqref{eigenv} are negative, the fully synchronous cluster
is stable. The instability occurs when either one real eigenvalue becomes positive,
or a pair of complex eigenvalues crosses the imaginary axis. The situation when 
one real eigenvalue  $\lambda_1$ changes from negative to positive value 
is exactly the transition described in 
section~\ref{sec:soq} above. 
One can check that the condition $\lambda_1=0$ in \eqref{eigenv}
in the limit of small coupling terms $\e_{1,2},\eta_{1,2}$ is exactly the 
condition $\cos\beta(1,\e_{1,2},\eta_{1,2})=0$ where $\beta$ is defined 
according to  \eqref{eq:ebeta}.

The comprehensive analysis of Eqs.~(\ref{e2},\ref{eigenv}) is hardly 
feasible due to a large number
of parameters. Therefore, we consider here below only a special case of 
isochronous oscillators, $\alpha=0$, which demonstrates both types 
of synchrony-breaking transition, of our interest in this study. Additionally, we
fix $\eta_2=0$, i.e. take purely dissipative nonlinear coupling.
Furthermore, we consider $\e_{1,2}\ge 0$.
Then, with account of Eq.~(\ref{e1_1}), we find
$\w=\w_0-\W=-\e_2$, what yields
\begin{equation}
\lambda_{1,2}=(1-2r^2)\pm\sqrt{r^4-\e_2^2}\;.
\end{equation}
\paragraph{Case of real eigenvalues.}
The condition for the eigenvalues to be real is $r^2\ge \e_2$. 
The bifurcation takes place when $\lambda_1$  
becomes zero, what yields
$r^2=\frac{1}{3}\left ( 2\pm\sqrt{1-3\e_2^2}\right )\ge \e_2$.
Hence, we have $\e_2\le 1/\sqrt{3}\approx 0.577$ and the critical line is found 
from the equation
\begin{equation}
2\pm\sqrt{1-3\e_2^2}=3\frac{1+\e_1}{1+\eta_1} \;.
\label{realroots}
\end{equation}
\paragraph{Bunch states.} Consider the case $\e_2=0$. 
The eigenvalues are $\lambda_1=1-r^2$, $\lambda_2=1-3r^2$. 
Hence, synchrony becomes unstable for 
$r<1$, i.e. for $\eta_1>\e_1$. Obviously, a neutrally stable state, 
$r=1$, $\W=\w_0$, and $\rho=R=\sqrt{\e_1/\eta_1}$ is a solution of 
Eq.~(\ref{e1}).
The case corresponds to the bunch state, cf. also \cite{PhysRevLett.94.164102}.  
\paragraph{Case of complex eigenvalues.}
The condition for the eigenvalues 
to be complex is $r^2 < \e_2$ and 
the condition for the real part 
to be zero is $r^2=\frac{1+\e_1}{1+\eta_1}=\frac{1}{2}$. 
Hence, the critical line is determined by $\eta_1=1+2\e_1$ and  $\e_2>0.5$.
For $\e_2=0.5$ and $\eta_1=1+2\e_1$ we have the ``Takens-Bogdanov point'' $\lambda_{1,2}=0$.
\paragraph{Stability of the asynchronous state.} This is accomplished as described in 
section \ref{wcl}. For the chosen parameters we obtain
from \eqref{eq:ebeta}
${\cal E}^2=\e_1^2+\e_2^2$ and $\beta=\arctan{(\e_2/\e_1)}$. Since $\e_{1,2}>0$, 
the asynchronous state is always unstable.

We emphasize that although synchrony breaking is quantified 
by only two eigenvalues~\eqref{eigenv}, the transition cannot be 
described as a low-dimensional bifurcation, 
because all oscillator leave the synchronous cluster simultaneously, 
what means that in the original $N$-dimensional phase
space there is $N-1$-fold degeneracy of eigenvalues~\eqref{eigenv}.

\begin{figure}[ht!]
\centerline{\includegraphics[width=0.48\textwidth]{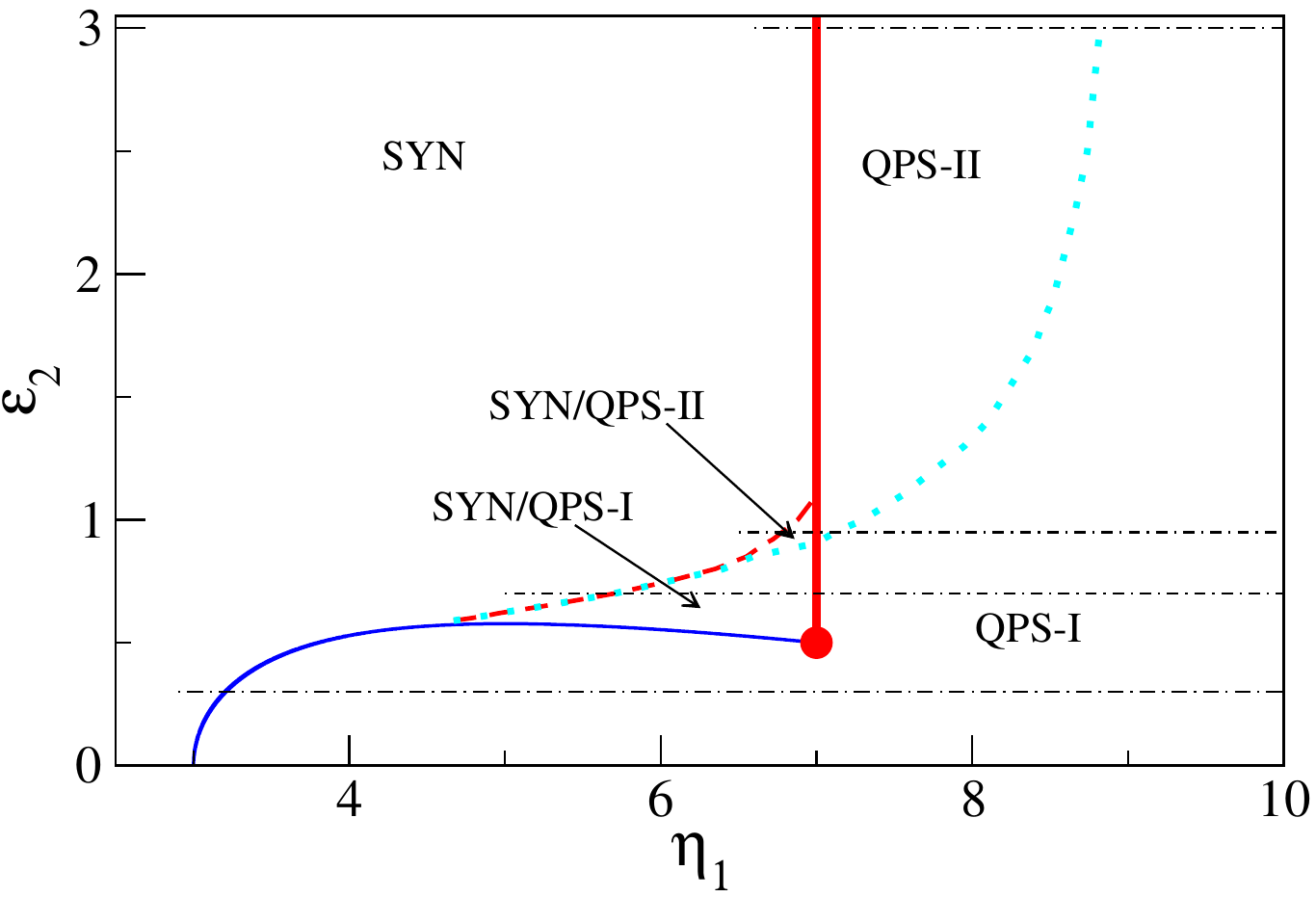}}
\caption{(Color online) Bifurcation diagram for $\e_1=3$. 
Blue solid line: here the largest real eigenvalue equals zero;
red bold line: here real parts of complex roots are equal to zero.
Red filled circle marks the Takens-Bogdanov point.
Theoretical analysis is complemented by the numerical study, which reveals 5 domains: 
stable full synchrony (SYN), quasiperiodic partial 
synchrony (QPS-I and QPS-II), and two domains where bistability between
synchrony and partial synchrony is observed (SYN/QPS-I and SYN/QPS-II).
Red dashed and cyan dotted curves are obtained numerically; they 
denote emergence of the amplitude modulation and of the frequency 
difference, respectively, see text for details.
The line $\e_2=0$, $\eta_1>3$ corresponds to the partially synchronous 
but not quasiperiodic bunch states.
Four horizontal black dashed-dotted lines show the cuts of the diagram
illustrated in 
Figs.~\ref{freqdep03},\ref{freqdep3},\ref{hyst07},\ref{hyst095} and in the text.
}
\label{bfdiag}
\end{figure}

For an example of the bifurcation diagram we fix $\e_1=3$; thus, our bifurcation 
parameters are $\e_2$ and $\eta_1$. The results are shown in Fig.~\ref{bfdiag}.
The blue solid line corresponds to Eq.~(\ref{realroots}); here the largest 
real eigenvalue turns zero. The red bold line $\eta_1=7$, $\e_2>0.5$ shows
where the Hopf-like bifurcation (complex eigenvalues)
takes place. 
Below the blue solid line and to the right of the red bold one the full 
synchrony is unstable and partially synchronous dynamics sets in. 
Next, we complement the diagram by the results of 
direct numerical simulation.

\section{Numerical exploration}
\label{sec:num}
All computations have been performed for $\w_0=5$ and $N=501$.
For several points in the diagram we checked that increasing of the ensemble
size up to several thousands does not influence the results.
We analyze the bifurcation diagram in Fig.~\ref{bfdiag}, 
by describing transitions at several fixed values of $\e_2$ (marked with dashed-dotted lines)
while parameter $\eta_1$ increases.

\begin{figure}[ht!]
\centerline{\includegraphics[width=0.48\textwidth]{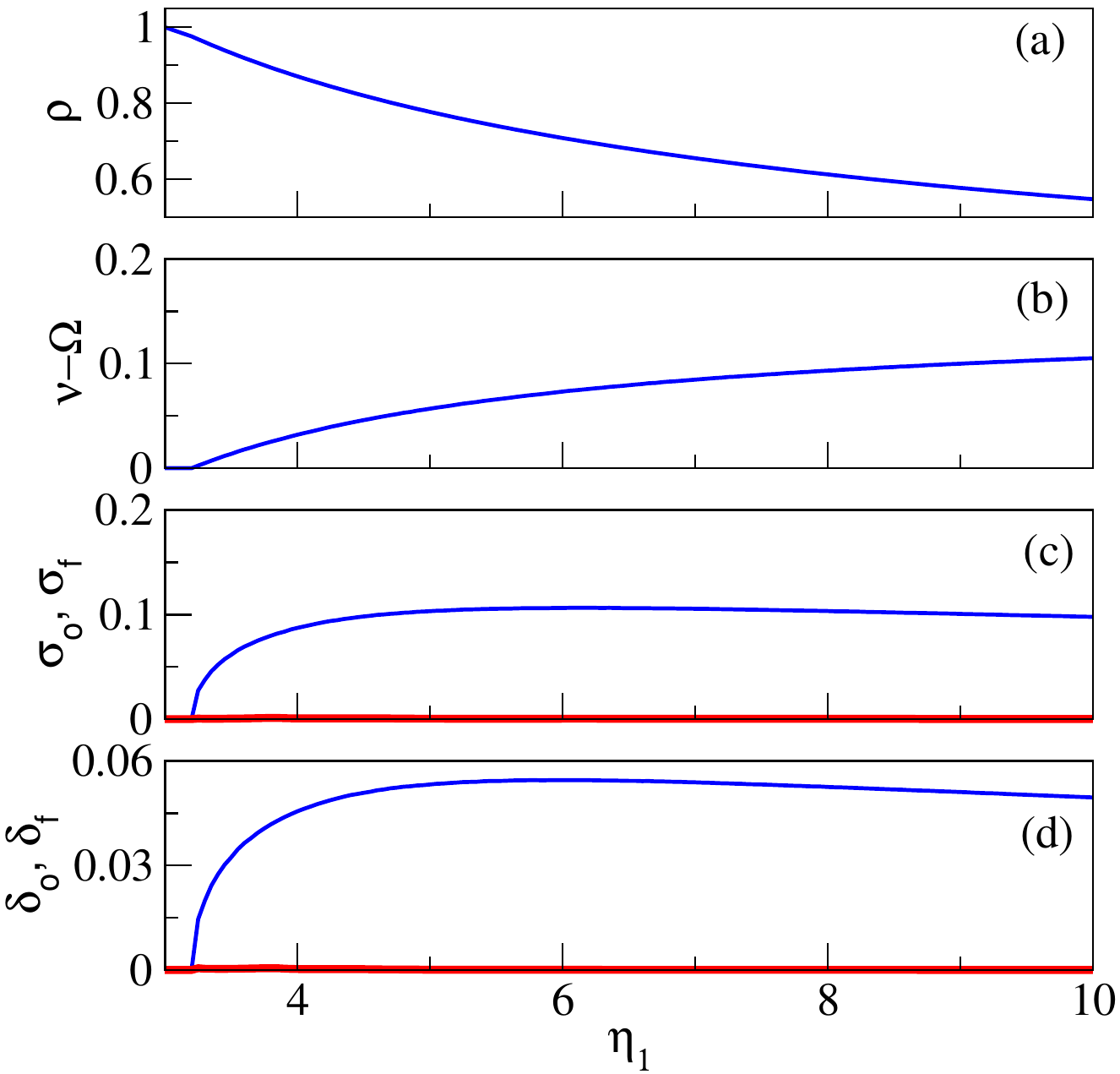}}
\caption{(Color online)
Transition from synchrony to quasiperiodic partial synchrony of type I
for small $\e_2=0.3$.
(a) Mean field amplitude $\rho$. (b) Frequency difference is the essential 
feature of the QPS-I dynamics. Here we see that the mean field is faster than the 
oscillators, and that the frequency difference $\nu-\W$ smoothly grows beyond the
transition point. 
(c) Frequency modulation of the oscillators and of the mean field is quantified
by the standard deviations of their instantaneous frequencies, denoted as
$\sigma_o$ and $\sigma_f$; 
these quantities are shown by blue solid and red bold curves, respectively. 
(d) Here blue solid and red bold curves show the standard deviations 
of oscillator and mean field amplitudes, denoted as $\delta_o$ and $\delta_f$,
respectively.
}
\label{freqdep03}
\end{figure}

\begin{figure}[ht!]
\centerline{\includegraphics[width=0.48\textwidth]{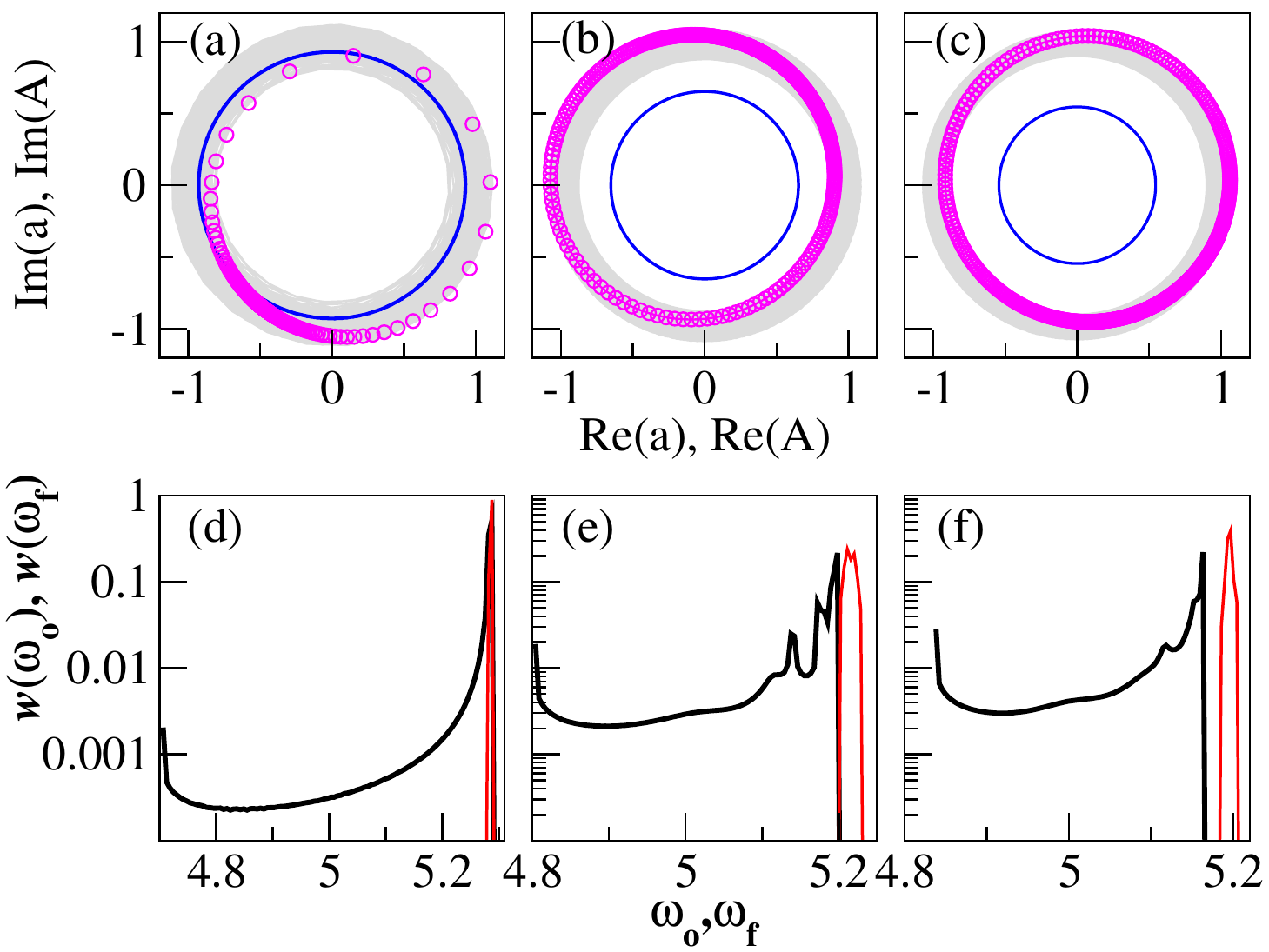}}
\caption{(Color online)
(a,b,c): Phase portraits, illustrating the synchrony-to-QPS-I
transition at $\e_2=0.3$. 
Gray solid and blue bold curves show trajectories of an oscillator
and of the mean field, respectively for $\eta_1=3.5$ (a), $\eta_1=7$ (b),
and $\eta_1=10$ (c). 
Open circles (magenta) show a snapshot of the ensemble. 
(d,e,f): Probability distribution $\rho$ of instantaneous frequency
of the oscillator, $\w_o$ (bold black curve), and of the mean field, 
$\w_f$ (red curve) for $\eta_1=3.5$, $\eta_1=7$,
and $\eta_1=10$, respectively.
}
\label{phplots03}
\end{figure}

For $\e_2=0, \eta_1>3$, the solution is a partially synchronous 
bunch state (not shown).
For small positive $\e_2$ (we have taken $\e_2=0.3$ for illustration), 
the dynamics beyond synchrony-breaking is quasiperiodic, 
as is shown in Fig.~\ref{freqdep03},\ref{phplots03} and corresponds to 
regime QPS-I as described
in section~\ref{sec:soq}.
Beyond the bifurcation, the frequency difference $\nu-\W$ (we remind that $\W$ and  
$\nu$ 
are frequencies of an oscillator and of the mean field, respectively) smoothly grows, as well 
as the amplitude modulation of oscillators (this can be also appreciated from 
the phase portraits in Fig.~\ref{phplots03}). 
It can be also recognized, that the distribution of the points in a snapshot
becomes more uniform with increase of $\eta_1$, what corresponds to decrease 
of the mean field amplitude. 
Notice that variations of the mean field frequency and of the amplitude 
are small, so that the mean field can be considered as harmonic.   

\begin{figure}[ht!]
\centerline{\includegraphics[width=0.48\textwidth]{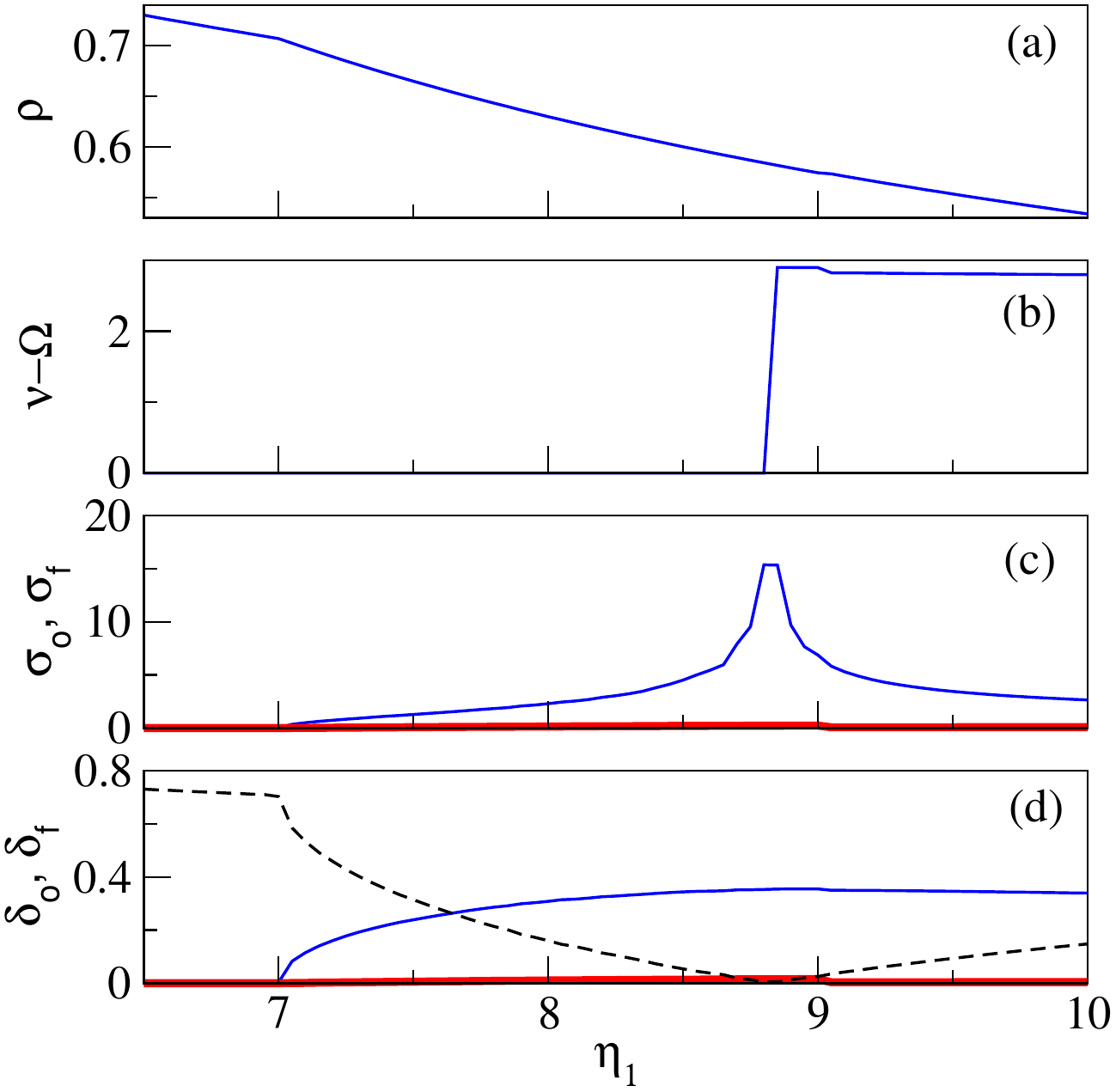}}
\caption{(Color online)
Transition from synchrony to QPS-II and then to QPS-I, for $\e_2=3$.
The shown quantities are same as in Fig.~\ref{freqdep03}.
In (d) we additionally show the minimal value of the oscillator 
amplitude $r$ (dashed curve).
The first transition, synchrony to QPS-II, takes place at 
$\eta_1=7$, as predicted by the stability analysis. 
At this point the amplitude and frequency modulation of 
oscillators emerge smoothly, but the
average frequencies are still equal, $\nu=\W$.
At $\eta_1\approx 8.8$ the second transition takes place,
here the frequency difference appears by jump.
}
\label{freqdep3}
\end{figure}

\begin{figure}[ht!]
\centerline{\includegraphics[width=0.48\textwidth]{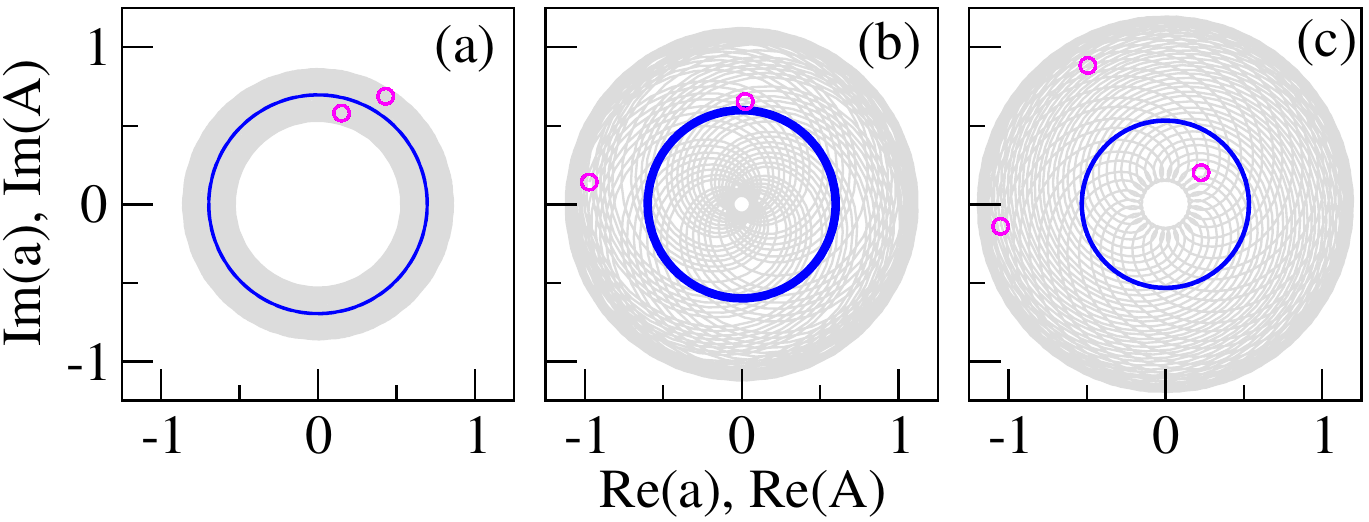}}
\caption{(Color online)
Phase portraits, illustrating the transitions at $\e_2=3$,
(the same curves as in Fig.~\ref{phplots03}). (a) QPS-II state, $\eta_1=7.1$.
(b) Close to the transition from QPS-II to QPS-I, $\eta_1=8.5$. 
(c) QPS-I state at $\eta_1=10$.
}
\label{phplots3}
\end{figure}

Next we consider large $\e_2=3$, see Fig.~\ref{freqdep3},\ref{phplots3}.
In contrast to the case of small $\e_2$, we see that, with increase of $\eta_1$,
the quasiperiodic motion initially appears due to the pure 
amplitude modulation (regime QPS-II). 
It means, that trajectory in the phase space lays on a torus that encircles the 
original limit cycle, and whose ``thickness'' grows smoothly with 
$\eta_1$. 
The ensemble elements split into several (quasi)clusters that rotate around the torus 
in such a way that $\nu=\W$.
(We checked that for the parameters used for the phase portrait plots, the 
averaged frequencies and the amplitudes of all elements are the same up 
to numerical precision).
Then, when $\eta_1$ attains some critical value, the frequency difference 
appears by a jump (dotted cyan curve in Fig.~\ref{bfdiag}), 
and we observe a transition from QPS-II to QPS-I. 
Geometrically, this transition can be described as follows. 
With increase of $\eta_1$, the torus becomes more and more ``thick'' so that 
the minimal oscillation amplitude $|a|$ decreases and reaches zero at some value of $\eta_1$. 
From now on the rotation of the cluster encircles the origin on the $x,y$ plane, 
and, hence, the frequencies of an oscillator 
and of the mean field start to differ. 
With further increase of $\eta_1$, the minimal $|a|$ grows, but the trajectory
continues to encircle the origin. 
Notice that at the transition from QPS-II to QPS-I,
the frequency difference $\nu-\W$ emerges by a jump and 
then remains practically constant. 


\begin{figure}[ht!]
\centerline{\includegraphics[width=0.48\textwidth]{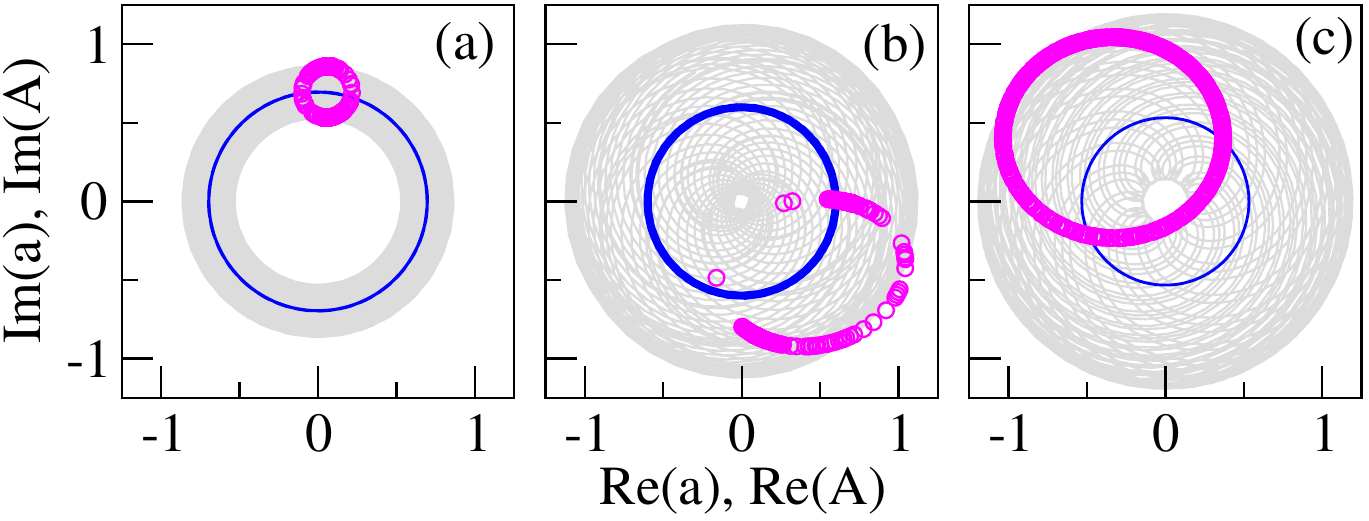}}
\caption{(Color online) Same as in Fig.~\ref{phplots3}, but for slightly 
non-identical oscillators. Now it is easy to see that the trajectory (represented
by a sequence of oscillators' states marked with circles) in (a) does 
not encircle the origin directly, but encircles the mean field, 
and therefore the frequency difference $\nu-\W=0$,
this is the QPS-I state. The trajectory in (c) directly encircles the origin and 
therefore the frequencies differ, this is the QPS-II state.
}
\label{phplots3nonid}
\end{figure}

In fact, the trajectories of (quasi)clusters cannot be easily recognized from 
the phase plots in Fig.~\ref{phplots3}. However, the dynamics becomes much more 
illustrative for slightly nonidentical units, as shown in Fig.~\ref{phplots3nonid}.
In this computation we take oscillator frequencies uniformly distributed in
$\w_0-\Delta,\w_0+\Delta$, where $\Delta=0.001$.  
Noteworthy, for small $\e_2$,  inhomogeneity does 
not affect the overall picture, but just slightly changes the threshold for
synchrony breaking.

\begin{figure}[ht!]
\centerline{\includegraphics[width=0.48\textwidth]{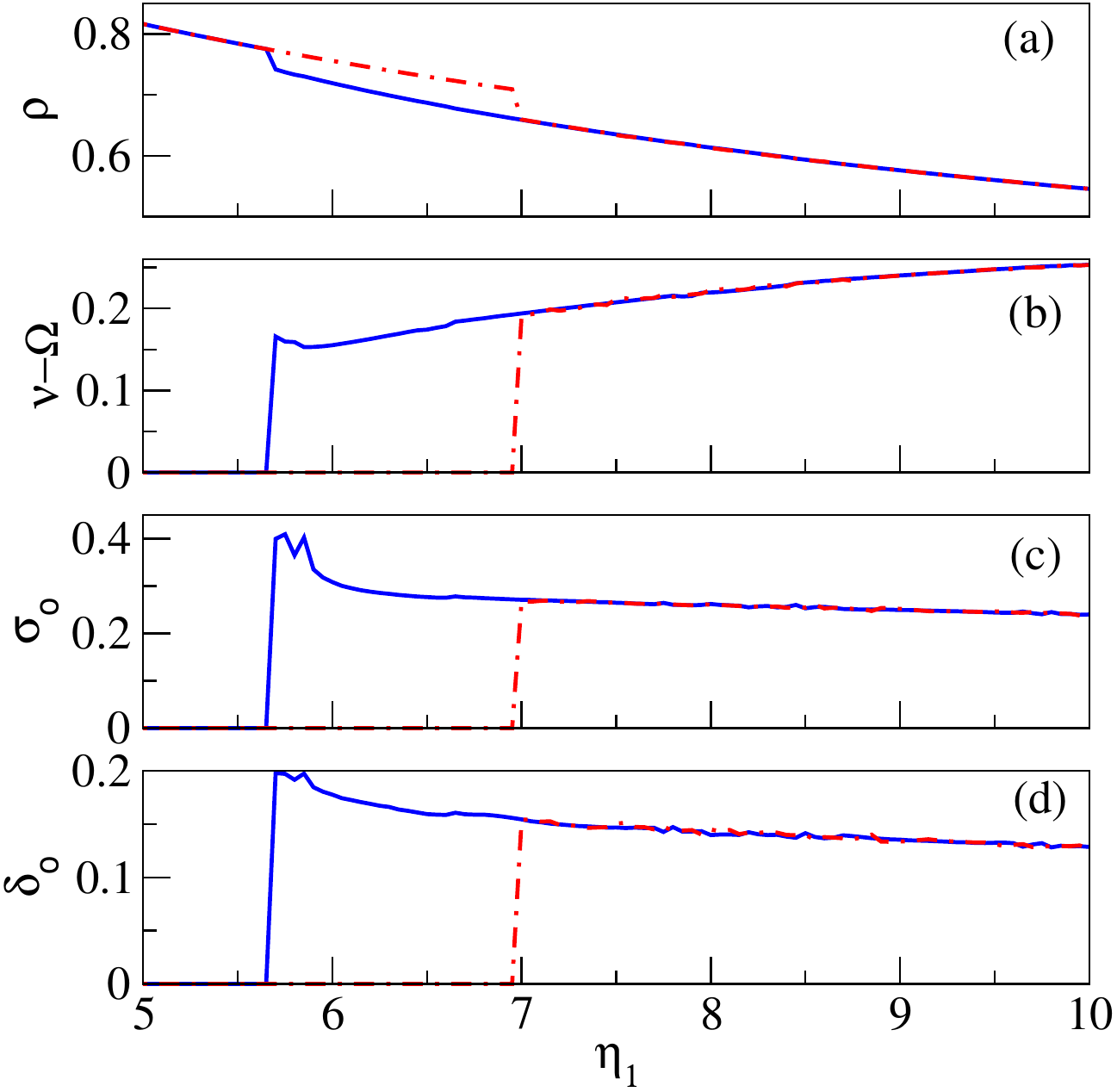}}
\caption{(Color online)
Hysteresis at the transition from synchrony to QPS-I for $\e_2=0.7$.
The shown quantities are mean field amplitude $\rho$ (a),
frequency difference $\nu-\W$ (b), and standard deviations $\sigma_o$ (c) 
and $\delta_o$ (d) of the instantaneous frequency and of the amplitude 
of the oscillators.
In each panel solid blue line shows the results obtained for nearly 
asynchronous initial conditions, while dashed-dotted red line 
corresponds to nearly synchronous initial conditions.
}
\label{hyst07}
\end{figure}

Finally, we consider the intermediate values of $\e_2$. The essential novel feature 
here is bistability and hysteresis.
It turns out that, with increase of $\eta_1$, a QPS state gains its stability 
while the synchronous state is still stable. 
Thus, partial and full synchrony coexist in this domain. 
(Practically, we performed simulations either starting from almost synchronous 
or from almost asynchronous initial conditions. 
Alternatively, to determine the stability domain of QPS we started 
from the partially synchronous state and decreased $\eta_1$.)
Analysis shows that this domain contains sub-domains of QPS-I and QPS-II dynamics.  
For illustration, we consider synchrony-breaking transitions for 
two values of parameter $\e_2$, $\e_2=0.7$ and $\e_2=0.95$. 
In the former case, we observed a transition from synchrony to the QPS-I dynamics.
In contrast to small $\e_2$ (cf. the picture for $\e_2=0.3$ in Fig.~\ref{freqdep03}), 
here the frequency 
difference $\nu-\W$ and the amplitude modulation appear by a jump, 
see Fig.~\ref{hyst07}. 
In the latter case, $\e_2=0.95$, we first observed a transition from synchrony 
to QPS-II and then another transition to QPS-I. 
It means that in this case, with increase of $\eta_1$,
first the amplitude modulation appears by a jump, and then the frequency difference appears at another
critical value of the parameter.
\begin{figure}[ht!]
\centerline{\includegraphics[width=0.48\textwidth]{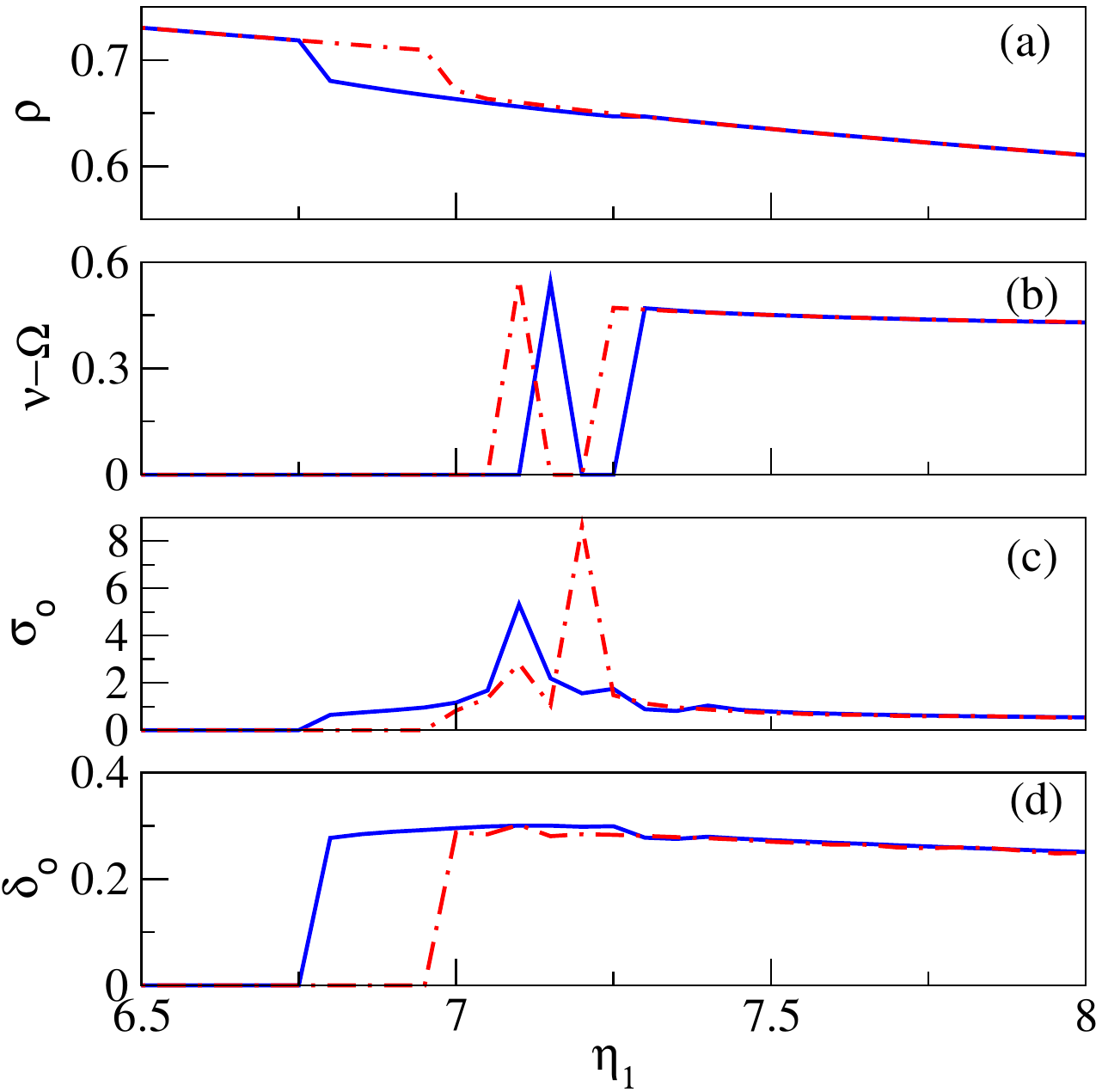}}
\caption{(Color online)
Hysteresis at transitions between synchrony, QPS-I and QPS-II for $\e_2=0.95$.
The shown quantities are the same as in Fig.~\ref{hyst07}.
}
\label{hyst095}
\end{figure}

To conclude this section, we mention that we cannot claim that the diagram 
in Fig.~\ref{bfdiag} yields a complete description of the dynamics, because 
it is not possible to check all possible initial conditions. 
For example, for some small parameter domain 
we have observed coexistence of both types of QPS dynamics. We cannot exclude 
other interesting dynamical regimes, but we believe that we have described the 
dominating solutions. 

\section{Discussion and conclusions}
\label{SumDis}
In this paper we analyzed two regimes of partially synchronous states in globally
coupled identical oscillator populations. These regimes can be attributed to a type
of bifurcation at the transition from full to partial synchrony: QPS-I corresponds to
one real evaporation eigenvalue becoming positive, while regime QPS-II corresponds
to two complex evaporation eigenvalues crossing real axis (in terms of multipliers, 
one real multiplier becomes larger than one or two complex multipliers become larger than
one in absolute values). These transitions are 
roughly related to two typical bifurcations from a steady state to a periodic dynamics:
SNIPER (saddle-node-infinite-period) and Hopf bifurcations. The main difference is that
in ensembles of oscillators the transition is extremely high-dimensional (in fact,
infinite-dimensional in the thermodynamic limit) so that usual 
low-dimensional bifurcation theory does not apply. In particular, while we can reliably
describe  some robust dynamical features like frequency difference between the mean field
and the individual oscillators, other fine dynamical features such as 
appearance of clusters seem to be non-universal and strongly model-dependent. 

For example, the simplest setup for the description of the regime QPS-I is 
the nonlinear extension of the Kuramoto-Sakaguchi
model  Eq.~(\ref{e1_2}), but it does not allow for
multiple clusters.
For phase models with a general coupling function (not just one harmonics)
of phase differences,
called Daido models \cite{Daido-93,*Daido-93a,*Daido-95,*Daido-96}, 
this does not hold. Therefore, for such models one can expect (i) 
scattered states, (ii) clustered states, (iii) and mixed states 
(scattered oscillators plus cluster(s)). 
These states may also be quasiperiodic. Certainly, the same can 
be said if one goes beyond the phase dynamics approximation and analyzes globally 
coupled multidimensional oscillators. 
To the type (iii) belong also
 chimera states
\citep{Kuramoto-Battogtokh-02,*Abrams-Strogatz-04,*PhysRevLett.100.044105,%
*Abrams-Mirollo-Strogatz-Wiley-08,*Martens2013}, originally described for 
non-locally coupled oscillators and for interacting subpopulations.  
In a chimera-like state of a globally coupled ensemble, one cluster coexists with
a scattered sub-population. 
This regime can be considered as a special case of partial synchrony;
recently studied examples include ensembles of phase oscillators with  delay 
\cite{Yeldesbay-Pikovsky-Rosenblum-14} and 
ensembles of  SL 
systems~\cite{Sethia-Sen-14,Schmidt_etal-14,*PhysRevLett.114.034101}.

Noteworthy, that chimera state was recently found \cite{Sethia-Sen-14}
in a well studied model of linearly coupled SL oscillators, 
see~\cite{PhysRevA.46.R7347,Nakagawa-Kuramoto-93,*Nakagawa-Kuramoto-94}.
The equations for complex variables $a_k$ read (in our notation): 
\begin{equation}
\dot a_k=(1+\ii\w_0)a_k-(1+\ii\alpha)|a_k|^2a_k+\bar\e(1+\ii \mu) (A-a_k) \;,
\label{nk1}
\end{equation}
where $A$ is defined according to Eq.~(\ref{nk3}).
Notice that in the weak-coupling approximation, 
this models reduces to the Kuramoto-Sakaguchi case with 
$\w=\w_0-\alpha-\e\sin\beta$, $\beta=\arctan\mu$, and $\e=\bar\e/\cos\beta$.
Thus, no partially synchronous state can be found for weak coupling.

If one goes beyond the phase approximation and considers the full equations, 
then, as shown in Ref.~\cite{Nakagawa-Kuramoto-93,*Nakagawa-Kuramoto-94}, 
there exists a parameter domain where 
both synchrony and full incoherence are unstable and therefore some partially
synchronous state appears.  
Namely, synchronous solution $a_k=A$ is always stable 
if $\alpha\mu> -1$. 
Otherwise, for $\alpha\mu < -1$,  synchrony is 
stable if $\e>\e_c=-\ds\frac{2(1+\alpha\mu)}{1+\mu^2}$.
Thus, with increase of 
coupling we can observe a transition incoherence - intermediate state - synchrony;
numerics shows that in the intermediate state the collective mode is chaotic
or exhibits a chimera state \cite{Sethia-Sen-14}.
Notice that the transition from the asynchronous state to the stable 
synchronous state happens when one real eigenvalue becomes negative.

\begin{figure}[ht!]
\centerline{\includegraphics[width=0.48\textwidth]{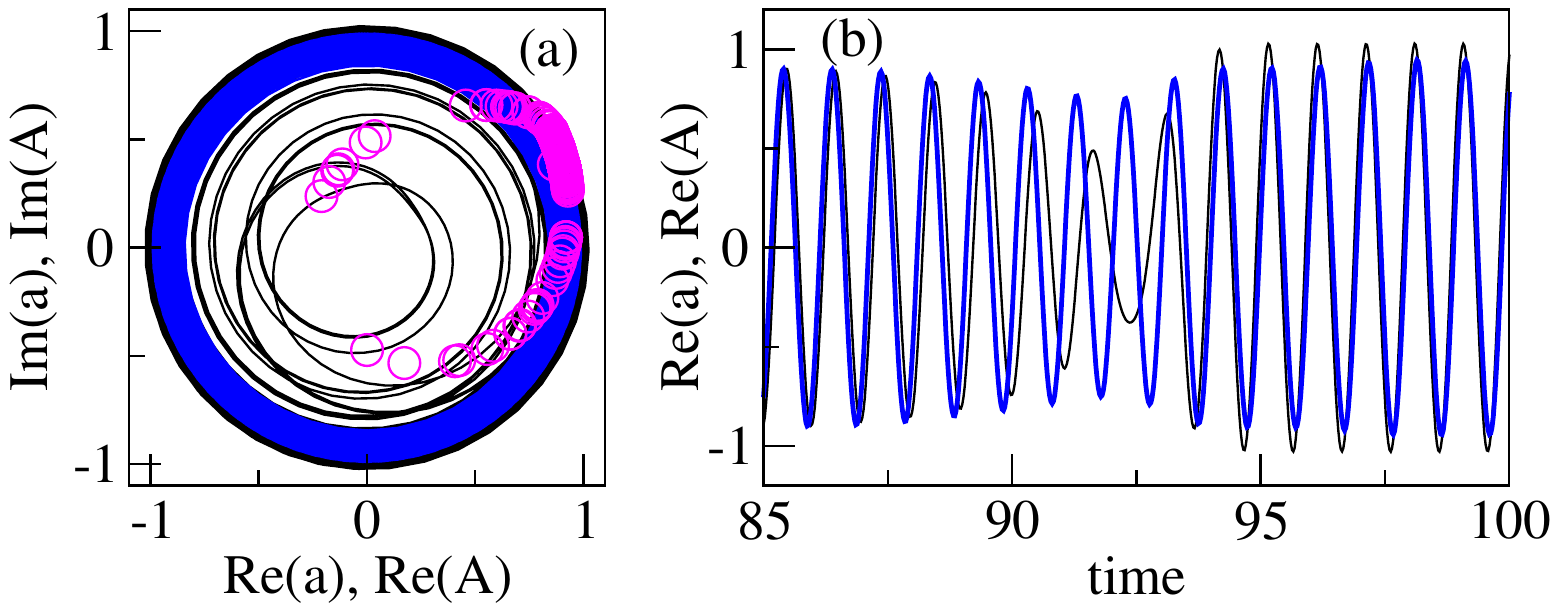}}
\caption{(Color online) Chaotic dynamics of the model \eqref{nk1}.
Black solid line shows trajectory of an oscillator; blue bold line
depicts the mean field. Open circles show a snapshot of the ensemble.
(a): phase portraits, (b): time dependence.
Notice then when the oscillator amplitude decreases, the mean field
makes an additional rotation with respect to the oscillator; thus,
the time-averaged frequencies differ, as is typical for a QPS-I state.
}
\label{fig:NK1}
\end{figure}

Remarkably,  in case of chaos the dynamics 
also possesses the property, characteristic of the QPS-I regime: 
the frequencies of the mean field and of the units are different.
For an example we take  the model~\eqref{nk1} 
with $\e=0.39$, $\alpha=-1.5$, $\mu=1$. For these parameter values the dynamics 
of both oscillators and of the collective mode is chaotic, see Fig.~\ref{fig:NK1}a.
Computation of average frequencies $\nu$ and $\W$ shows that $\nu\ne\W$.
This fact obviously follows from the plots of $\mbox{Re}[a(t)]$, 
$\mbox{Re}[A(t)]$, see Fig.~\ref{fig:NK1}b:
when the amplitude of an oscillator becomes relatively small due to chaos, 
the phase slip occurs because the mean field makes an additional rotation with respect to 
the oscillator. This picture agrees with a qualitative description of
phase synchronization of chaos \cite{Rosenblum-Pikovsky-Kurths-96}, where 
the effect of chaotic amplitudes is considered as an effective noise which causes
phase slips. Phase slips of chaotic oscillators with respect to the mean field 
can be also observed for intrinsically
chaotic systems, e.g., for a ensemble of globally coupled R\"ossler oscillators. 

Finally, we notice that quasiperiodic partially synchronous states can appear 
without synchrony-breaking transition. The most known example is the van Vreeswijk 
model \cite{vanVreeswijk-96,*Mohanty-Politi-06}
of leaky integrate-and-fire neurons where the quasiperiodic motion emerges
from the splay state. 

In summary, we have analyzed a model of nonlinearly-coupled 
limit-cycle oscillators and revealed two routes to two quasiperiodic 
states via synchrony breaking. These states appear via two 
different bifurcations. 
Moreover, we have shown the transition between QPS-I and QPS-II states, 
as well as domains of bistability. 

\acknowledgments
The study was supported by COSMOS ITN (EU Horizon 2020 research and innovation programme
under Maria-Sklodowska-Curie grant agreement No 642563)
We acknowledge helpful discussions with A. Politi and M. Zaks.
A. P. was supported by 
the grant (agreement 02.B.49.21.0003 of August 27, 2013  between the 
Russian Ministry of Education and Science and Lobachevsky State University of
Nizhni Novgorod).

%

\end{document}